\documentclass[preprint,12pt]{elsarticle}

\usepackage{graphicx}
\usepackage{dcolumn}
\usepackage{bm}
\usepackage{amssymb}
\usepackage{tabularx}

\usepackage[margin=1in]{geometry}

\usepackage{setspace}

\biboptions{sort}

\journal{Journal of Solid State Chemistry}

\begin{document}
\begin{frontmatter}

\title{Evolution of Magnetism in the Na$_{3-\delta}$(Na$_{1-x}$Mg$_x$)Ir$_2$O$_6$ Series of Honeycomb Iridates}

\author[jhuchem,iqm]{David C. Wallace}

\author[NCNR,UDel]{Craig M. Brown}

\author[jhuchem,iqm]{Tyrel M. McQueen}
\ead{mcqueen@jhu.edu}

\address[jhuchem]{Department of Chemistry, The Johns Hopkins University, Baltimore, MD 21218, USA}

\address[iqm]{Institute for Quantum Matter, Department of Physics and Astronomy, The Johns Hopkins University, Baltimore, MD 21218, USA}

\address[NCNR]{NIST Center For Neutron Research, Gaithersburg, MD, 20899, USA}

\address[UDel]{Department of Chemical and Biomolecular Engineering, University of Delaware, Newark, Delaware 19716, USA}

\date{\today}
\begin{abstract}
The structural and magnetic properties of a new series of iridium-based honeycomb lattices with the formula Na$_{3-\delta}$(Na$_{1-x}$Mg$_x$)Ir$_2$O$_6$ (0~$\leq$~{\it x}~$\leq$~1) are reported. As {\it x} and $\delta$ are increased, the honeycomb lattice contracts and the strength of the antiferromagnetic interactions decreases systematically due to a reduction in Ir--O--Ir bond angles. Samples with imperfect stoichiometry exhibit disordered magnetic freezing at temperatures {\it T$_f$} between 3.4 K and 5 K. This glassy magnetism likely arises due to the presence of non-magnetic Ir$^{3+}$, which are distributed randomly throughout the lattice, with a possible additional contribution from stacking faults. Together, these results demonstrate that chemical defects and non-stoichiometry have a significant effect on the magnetism of compounds in the {\it A}$_2$IrO$_3$ materials family.
\end{abstract}

\end{frontmatter}

\section{Introduction}

5{\it d} transition metal oxide materials have recently attracted significant interest due to the comparable energy scales of crystal field stabilization, electronic correlations (Hubbard U), and spin-orbit coupling \cite{MottStrongSOC}. In particular, iridium oxides are of great interest due to the ability of Ir to adopt a {\it d}$^5$ electronic configuration in its 4+ oxidation state, yielding a singly-occupied orbital with non-trivial spin texture in the strong spin-orbit coupling limit. This electronic configuration has been predicted to give rise to an unusual pattern of superexchange interactions, resulting in magnetic frustration and possible spin liquid behavior \cite{Na2IrO3Prediction,XtalfieldandcorrelationA2IrO3,A2IrO3KitaevHeisenberg}. For this reason, the {\it A}$_2$IrO$_3$ ({\it A} = Li, Na) family of layered honeycomb materials have been the subject of intense scrutiny \cite{Kimchi,Andrade,Rau,Lei,Katukuri}. 

The majority of experimental work on Na$_2$IrO$_3$ and Li$_2$IrO$_3$ has focused on single-crystalline specimens grown via ``self-flux." Despite strong antiferromagnetic interactions in Na$_2$IrO$_3$, evidenced by a reported large negative Weiss temperature $\theta_W$ = -116 K, long-range antiferromagnetic order is only present below a N\'{e}el temperature of $T_N$ = 15 K, demonstrating that that Na$_2$IrO$_3$ is in a magnetically frustrated regime \cite{Na2IrO3Synthesis,Na2IrO3revisedstructure}. A study of the (Na$_{1-x}$Li$_x$)$_2$IrO$_3$ solid solution revealed that  $\theta_W$ as well as $T_N$ vary significantly with {\it x}, and the magnetism reaches a peak in frustration at the intermediate value of {\it x} = 0.7 \cite{NaLiDoping}. Isovalent substitution of Li for Na in the {\it A}$_2$IrO$_3$ structure presumably has the effect of modulating the exchange interaction strengths between neighboring Ir sites, however this effect has not been systematically explored. Recently, a broad variety of new ternary Ir oxides with the formula Na$_{3-\delta}${\it M}Ir$_2$O$_6$ ({\it M} = Zn, Cu, Ni, Co, Fe and Mn) were reported \cite{BaroudiCava}, but no unifying picture of the magnetism demonstrated by this family can yet be drawn. 

Here, we report another new member of this chemical family, Na$_{3-\delta}$(Na$_{1-x}$Mg$_x$)Ir$_2$O$_6$, in which Mg substitutes for Na in the honeycomb planes and the Na content between planes is variable (Fig.~\ref{structure}(a)). Aliovalent substitution of Mg for Na in the honeycomb plane offers a convenient and controllable method by which to tune the lattice parameters of the unit cell and, in doing so, vary the exchange interactions between neighboring Ir sites. Additionally, we resolve the discrepancy in magnetic behavior between polycrystalline and single-crystalline samples of Na$_2$IrO$_3$, where the former are spin glasses and the latter exhibit long range antiferromagnetic (AFM) order \cite{Na2IrO3Synthesis}, by showing that Na$_2$IrO$_3$ decomposes rapidly in air as previously reported, and that this decomposition destroys AFM order \cite{Na2IrO3hydration}. 

\section{Materials and Methods}
Polycrystalline Na$_2$IrO$_3$ was prepared via a method similar to what was previously reported \cite{Na2IrO3Synthesis}: Na$_2$CO$_3$ (NOAH Technologies Corp., 99.9\%)\footnote{Certain commercial suppliers are identified in this paper to foster understanding. Such identification does not imply recommendation or endorsement by the National Institute of Standards and Technology, nor does it imply that the materials or equipment identified are necessarily the best available for the purpose.}, MgO (NOAH Technologies Corp., 99.99\%) and Ir black (J\&J Materials Inc.) were intimately ground using an agate mortar and pestle and pelletized. The sample was heated in a covered alumina crucible to 750 $^{\circ}$C over a period of 4 hrs, held at that temperature for 30 hrs, then quenched in air. The pellet was then reground with an additional 5 mol \% Na$_2$CO$_3$, pelletized, and placed into a furnace preheated to 900 $^{\circ}$C where it dwelled for and additional 48 hrs. After the final heating, samples were removed from the furnace and allowed to cool in an argon filled glovebox to prevent reaction with laboratory air. Polycrystalline samples of Na$_{3-\delta}$(Na$_{1-x}$Mg$_x$Ir$_2$)O$_6$ were prepared via a similar method: Na$_2$CO$_3$, MgO and Ir metal powder were intimately ground and pelletized. Samples were heated in covered alumina crucibles to 750 $^{\circ}$C over a period of four hrs, held at that temperature for 30 hrs, then quenched in air. Samples were then reground, pelletized, and placed into a furnace preheated to 900 $^{\circ}$C where they dwelled for between 48 and 96 hrs, followed by quenching in air. The 900 $^{\circ}$C heating was repeated with intermediate regrindings until the laboratory x-ray powder diffraction (XRPD) data showed no impurity phases and no change in lattice parameters or relative peak intensities between subsequent heatings. After the final heating, samples were processed and stored in an argon-filled glovebox to prevent reaction with air. Samples of suitable size for neutron powder diffraction (NPD) were prepared via a similar method, but with an additional two week heating at 900 $^{\circ}$C.

Laboratory XRPD data were collected on a Bruker D8 Focus diffractometer using Cu-K$\alpha$ radiation with $\lambda_1$~=~1.5406   \AA~and $\lambda_2$~=~1.5445   \AA~and equipped with a LynxEye CCD detector. Rietveld refinements to laboratory XRPD data were performed using TOPAS (Bruker AXS). NPD data were collected  at the NIST Center for Neutron Research using the BT-1 high-resolution powder diffractometer with an incident wavelength of $\lambda$~=~1.5419   \AA~using a Cu (311) monochromator and with 15' and 60' for primary and secondary collimation, respectively. Rietveld refinements to NPD data were performed in GSAS/EXPGUI \cite{GSAS,EXPGUI}. Samples were loaded into vanadium cells and sealed in an inert helium glovebox to prevent exposure to air. DC and AC magnetization measurements were carried out using a Quantum Design Physical Properties Measurement System. Elemental analyses were performed via inductively-coupled plasma optical emission spectroscopy (ICP-OES) by Evans Analytical Group. XRPD simulations were performed using DIFFaX \cite{DIFFaX}.

\section{Results and Discussion}

\subsection{Structure}

\begin{figure}
\begin{centering}
\includegraphics[width=5.5in]{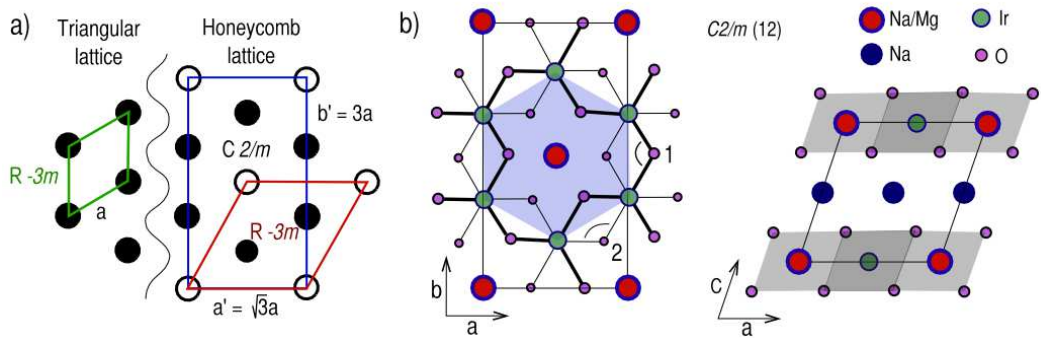}
\caption{{\bf (a)} A simple triangular lattice can be described by an R{\it-3m} unit cell (green) with in-plane lattice parameter {\it a}. Formation of honeycomb order by substitution of one third of the atoms in the triangular lattice is accommodated by a $\sqrt{3}$ expansion of the triangular lattice constant yielding a larger R{\it-3m} unit cell (red). Different stacking arrangements (not shown) can require a lower symmetry unit cell, such as the C2/{\it m} cell shown (blue). {\bf (b)} The structure of Na$_{3-\delta}$(Na$_{1-x}$Mg$_x$)Ir$_2$O$_6$ is described by the C2/{\it m} cell shown in the {\it ab}-plane. Oxygen atoms above the honeycomb plane are bonded to Ir with thick black lines, while those below the plane are bonded to Ir with thin lines. IrO$_6$ octahedra share edges to form a honeycomb lattice, highlighted by a light blue hexagon. Two distinct Ir--O--Ir bond angles, labeled 1 and 2, are possible in this structure. The structure is also shown in the {\it ac}-plane to highlight the stacking arrangement corresponding to the C2/{\it m} cell.}
\label{structure}
\end{centering}
\end{figure}

The structure of the {\it A}$_2$IrO$_3$ family of materials is derived from the layered triangular lattice compound $\alpha$--NaFeO$_2$ (Fig.~\ref{structure}(a)): when Na substitutes for one third of the Fe sites in NaFeO$_2$ in an ordered fashion, a layered honeycomb lattice is formed and the chemical formula becomes Na$_2$FeO$_3$. For this reason, the structural formula of {\it A}$_2$IrO$_3$ can be described more intuitively as {\it A}$_3${\it A}$'$Ir$_2$O$_6$, where electronically active {\it A}$'$Ir$_2$O$_6$ honeycomb layers, comprised of tilted, edge-sharing octahedra, are separated by electronically inert layers of {\it A} cations. The honeycomb order in these layers is energetically favorable due to the large differences in atomic radii and oxidation states of Ir in comparison to the {\it A}$'$ cation. Because electronically active honeycomb layers are separated by electronically inert triangular layers, the {\it A}$_2$IrO$_3$ family are essentially layered two dimensional materials. This fact can give rise both to interesting physics, and to complex structural disorder. 

\begin{figure}
\begin{centering}
 \includegraphics[width=3.5in]{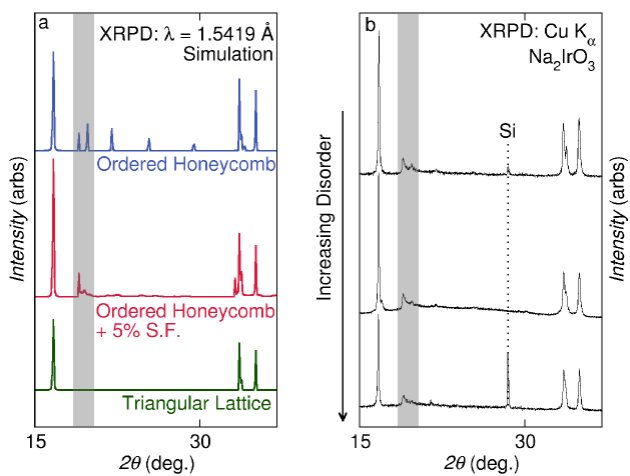}
 \caption{{\bf (a)} XRPD patterns simulated using DIFFaX are shown for layered structures of the following types: triangular (green), ordered honeycomb (blue), and ordered honeycomb with 5\% stacking faults (red). The influence of honeycomb ordering on the XRPD pattern is most apparent in the shaded region between 2$\theta$~=~18$^{\circ}$ and 22$^{\circ}$. {\bf (b)} XRPD data collected on three different samples of Na$_2$IrO$_3$ with varying degrees of structural disorder. The top sample has the fewest stacking faults, as illustrated by the sharp peaks in the shaded region between 2$\theta$~=~18$^{\circ}$ and 22$^{\circ}$.}
 \label{stacking}
 \end{centering}
 \end{figure}

Fig.~\ref{stacking}(a) shows simulated XRPD data that demonstrate how disordered stacking arrangements affect the observed XRPD pattern. When honeycomb lattices are stacked in a perfectly ordered manner, sharp reflections are observed in the XRPD data between scattering angle 2{\it $\theta$}~=~19$^{\circ}$ and 32$^{\circ}$ -- these are often referred to as supercell reflections, as the arise from the additional honeycomb ordering within the triangular layer. This structure (Fig.~\ref{structure}(a)) can be accounted for by a larger R$\overline{3}m$ cell where the {\it a} axis is expanded by $\sqrt{3}$ compared to the triangular R$\overline{3}m$ cell, or to lower symmetry variations such as the monoclinic C2/{\it m} or C2/{\it c} cells commonly observed (Fig.~\ref{structure}(b)) \cite{Na2IrO3revisedstructure,Cu5SbO6}. If stacking faults are introduced, even with a low probability of occurrence, the supercell reflections become broadened and attenuated as shown in Fig.~\ref{stacking}(a) with 5\% stacking faults. Fig.~\ref{stacking}(b) shows XRPD data collected on three different samples of Na$_ 2$IrO$_3$, each with a different degree of structural disorder. Rietveld refinement to data collected on a sample with stacking faults can incorrectly account for the observed loss in supercell reflection intensity by introducing cation site mixing in the honeycomb plane. The energy cost associated with producing a stacking fault in Na$_2$IrO$_3$ and structurally analogous systems is very small, and is thus significantly more likely to occur than antisite disorder \cite{Na2IrO3revisedstructure,Li2MnO3}. Similarly, a refinement can give the illusion of small particle size in order to account for the observed broadening of the supercell reflections. The same effect is observed in neutron diffraction, though the honeycomb order may be less apparent if atoms in the honeycomb layer have similar neutron scattering cross sections. It is therefore quite difficult to make accurate structural characterizations of samples with significant stacking disorder using neutron or x-ray diffraction techniques.

\begin{figure}
 \begin{centering}
\includegraphics[width=3.5in]{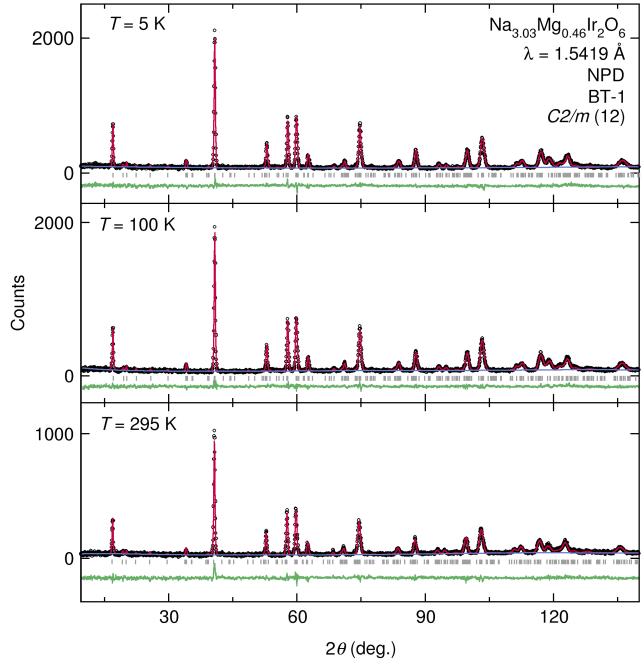}
 \caption{NPD data (black circles) collected on Na$_3$Mg$_{0.5}$Ir$_2$O$_6$ at {\it T} = 5 K, 100 K, and 295 K are shown along with Rietveld refinements (red) and the difference between the data and the fit (green). Tick marks (gray) indicate the positions of expected Bragg reflections.}
 \label{NPD671}
 \end{centering}
 \end{figure}

\begin{figure}
\begin{centering}
\includegraphics[width=3.5in]{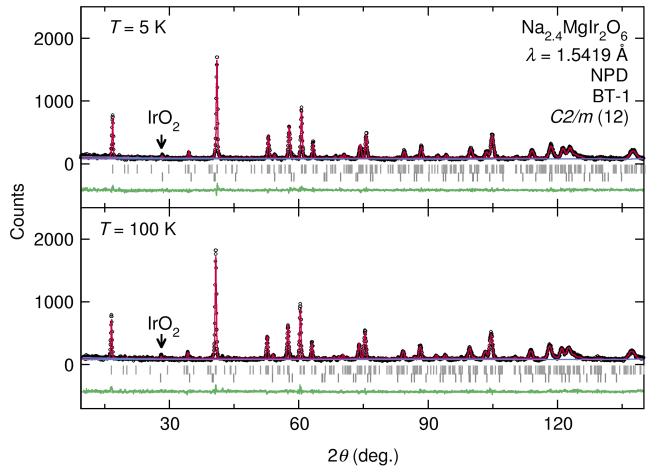}
\caption{NPD data (black circles) collected on Na$_{2.4}$MgIr$_2$IrO$_6$ at {\it T} = 5 K and 100 K are shown along with Rietveld refinements (red) and the difference between the data and the fit (green). A small ($\sim$ 10\%) IrO$_2$ impurity was present in the sample, and was included as a separate phase in the refinements. Tick marks (gray) indicate the positions of expected Bragg reflections. Tick marks corresponding to the IrO$_2$ impurity peak positions are located below the tick marks for the structure.}
\label{NPD672}
\end{centering}
\end{figure}

Fig.~\ref{NPD671} shows NPD datasets collected on Na$_{3}$Mg$_{0.5}$Ir$_2$O$_6$ at {\it T} = 5, 100, and 295 K, along with Rietveld refinements to the data. Small supercell reflections were observed that could not be indexed using the standard R$\overline{3}m$ cell for a triangular lattice. To account for this, a refinement was attempted using an R$\overline{3}m$ cell with a $\sqrt{3}$ expansion of the {\it a}-axis from the corresponding triangular cell, but this did not accurately index the observed supercell reflections. The reflections were properly indexed using the lower symmetry monoclinic space group C2/{\it m} (12). In order to determine cation site occupancies, it was assumed that no site mixing occurs between sites in the honeycomb plane (4{\it g} and 2{\it a}), and the 4{\it g} site was fixed at full Ir occupancy. The 2{\it a} site was assumed to have mixed Na/Mg occupancy. It was further assumed that Mg only substitutes for Na at the 2{\it a} site. The interplane Na sites (2{\it d} and 4{\it h}) were constrained to have the same occupancy, as were the O1 and O2 sites (8{\it j} and 4{\it i}). Occupancies were then allowed to refine freely within the constraints. The occupancies determined from refinement to the {\it T} = 5 K data were then fixed in refinements to data collected at higher temperatures. Due to the high correlation between isotropic displacement and occupancy, occupancies were determined with fixed {\it U$_{iso}$}~=~0.005, then {\it U$_{iso}$} was allowed to refine freely. The results of these refinements are listed in Table~\ref{NPDtable1}. Fig.~\ref{NPD672} shows NPD datasets collected on Na$_{2.4}$MgIr$_2$O$_6$ at {\it T} = 5 K and 100 K. Rietveld refinements to these datasets were carried out in the same way as for the previous sample, however, a small IrO$_2$ impurity was present in this sample, and was included as a separate phase in the refinements. The results of these refinements are listed in Table~\ref{NPDtable2}. Lattice parameters for Na$_{2.4}$MgIr$_2$O$_6$ at room temperature were determined from laboratory XRPD data collected using an air-free sample holder sealed under argon.

To confirm the compositions of the samples used for NPD, elemental analyses were performed via ICP-OES. The results of these analyses are shown in Table~\ref{ICPOES}. Initial attempts to dissolve the samples were unsuccessful, and the samples had to be recovered and re-dissolved using microwave digestion. While the Na/Mg ratios agree well with the compositions determined from NPD, the Ir content is anomalously low for both samples, which is likely due to the difficulty associated with fully dissolving Ir and its oxides. 

The structures of all other samples in the series were characterized via laboratory XRPD. Rietveld refinements were carried out using the reported structure for Na$_2$IrO$_3$ (space group C2/{\it m}) \cite{Na2IrO3revisedstructure}. However, because Ir is the dominant x-ray scatterer and stacking disorder is present, the {\it a} and {\it b} lattice parameters of the monoclinic cell were first obtained by performing a Le Bail fit to the data using space group R$\overline{3}m$, where the {\it a}-axis is the nearest-neighbor distance in the honeycomb lattice. The resulting {\it a} lattice parameter was converted to the the {\it a} and {\it b} lattice parameters of the monoclinic cell, which were then fixed for the remainder of the refinement. In this way, site mixing between the 4{\it g} and 2{\it a} sites can be introduced in order to improve relative peak intensities without adversely affecting the lattice parameters. This is particularly important when comparing compounds within a series that may have varying degrees of stacking disorder and site mixing. For reporting purposes, we refer to the resulting site-mixing percentages as the percent of disorder, as they are likely due to stacking faults rather than site mixing. The lattice constants and percents of disorder determined for these compounds are listed in Table~\ref{compoundstable}. 

\begin{figure}
\begin{centering}
\includegraphics[width=3.5in]{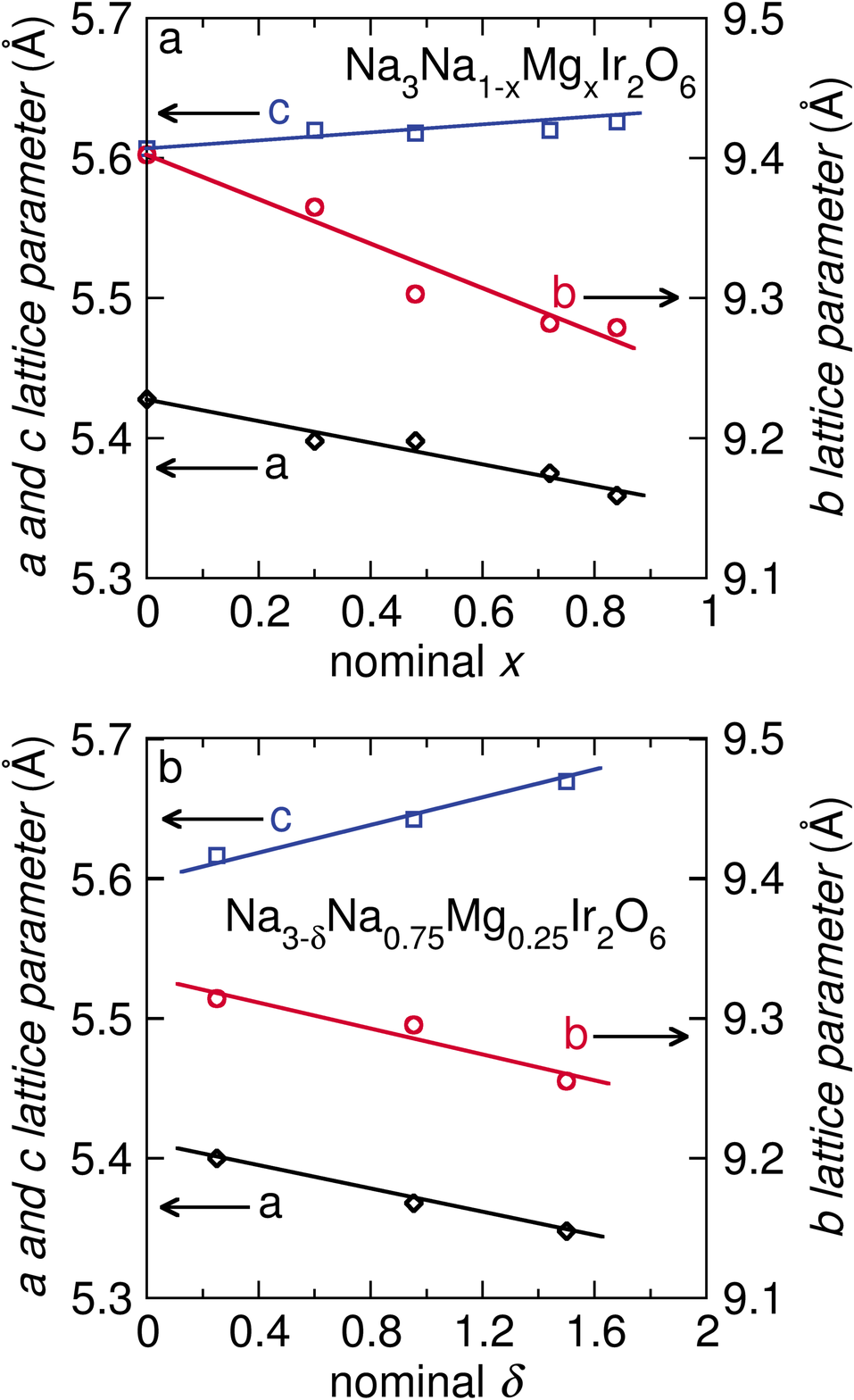}
\caption{Plots of lattice parameters vs. nominal chemical composition for the cases of varied Mg substitution (Na$_3$(Na$_{1-x}$Mg$_x$)Ir$_2$O$_6$) {\bf (a)} and varied Na-content (Na$_{3-\delta}$(Na$_{0.75}$Mg$_{0.25}$)Ir$_2$O$_6$) {\bf (b)}. Substitution of Mg for Na in the honeycomb plane causes the {\it a}- and {\it b}-axes (black diamonds and red circles, respectively) to contract and the {\it c}-axis (blue squares) to expand. Decreasing the amount of sodium in the lattice has a similar effect.}
\label{latticeparams}
\end{centering}
\end{figure}

In order to distinguish between the effects of Mg substitution and Na deficiency, two series of compounds were prepared using the exact same heating schedule. The first series were prepared with the nominal formula Na$_3$Na$_{1-x}$Mg$_x$Ir$_2$O$_6$, and the second series had the nominal formula Na$_{3-\delta}$Na$_{0.75}$Mg$_{0.25}$Ir$_2$O$_6$. Plots of lattice parameters versus nominal composition demonstrate the influence of both Mg substitution (Fig.~\ref{latticeparams}(a)) and Na-deficiency (Fig.~\ref{latticeparams}(b)) on the lattice parameters. Substitution of Mg for Na in the honeycomb plane causes the {\it a}- and {\it b}-axes to contract, and the {\it c}-axis to expand slightly. This result supports the assumption that Mg substitutes into the honeycomb plane, as substitution between the planes would cause the {\it c}-axis to contract due to electrostatics and the reduced size of Mg relative to Na. Similarly, Na-deficient samples show a contraction of the {\it a}- and {\it b}-axes and an expansion of the {\it c}-axis. Because both Na and Mg contents are variable, compounds in the Na$_{3-\delta}$(Na$_{1-x}$Mg$_x$)Ir$_2$O$_6$ series have a broad range of possible unit cell parameters.

\subsection{Magnetic Properties}

\begin{figure}
\begin{centering}
\includegraphics[width=3.5in]{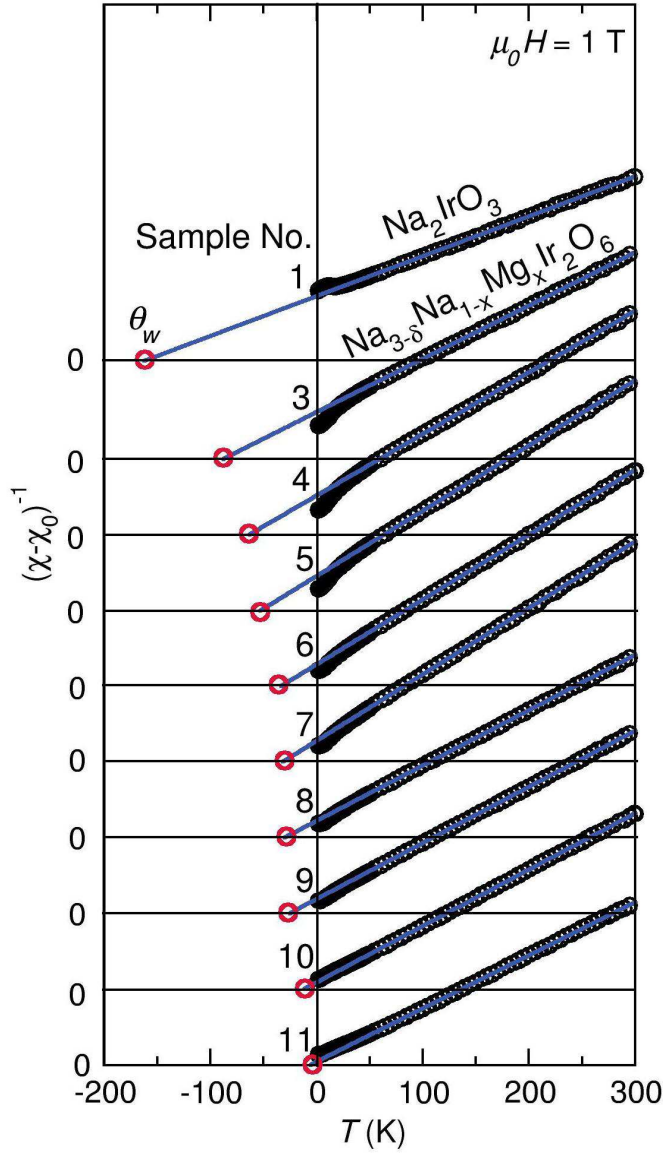}
\caption{Inverse magnetic susceptibility is plotted against temperature for each member of the series. Datasets are shifted along the y-axis for visibility. Linear fits to inverse magnetic susceptibility data over the range {\it T} = 60 K to 300 K yield the Curie constant C and the Weiss temperature $\theta_W$ for each member of the series. Datasets are shifted along the y-axis for visibility. The Weiss temperatures $\theta_W$ for each member of the series are indicated by red circles, corresponding to the x-intercept of the linear extrapolation from the Curie-Weiss regime.}
\label{CWanalysis}
\end{centering}
\end{figure}

\begin{figure}
\begin{centering}
\includegraphics[width=3.5in]{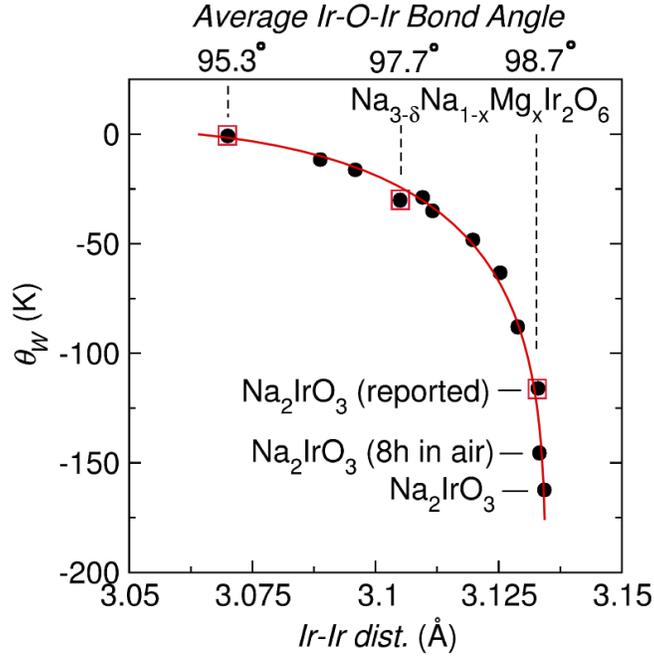}
\caption{Plot of the Weiss temperature $\theta_W$ versus the average nearest neighbor Ir--Ir distance for all compounds studied in this work. Negative values of $\theta_W$ indicate that antiferromagnetic interactions dominate the magnetic susceptibility of these layered honeycomb iridates, and the magnitude of $\theta_W$ is a measure of the strength of these interactions. Expansion of the honeycomb plane increases the overall antiferromagnetic interaction strength as Ir--O--Ir bond angles deviate from 90$^{\circ}$. Average Ir--O--Ir bond angles, determined from analysis of NPD data, are shown for two points on the plot, as well as the reported value for Na$_2$IrO$_3$, denoted by red boxes \cite{Na2IrO3revisedstructure}. $\theta_W$ is largest in Na$_2$IrO$_3$ and is diminished by chemical substitution and sodium vacancies, as well as decomposition in air. The smooth curve through the data is a guide to the eye.}
\label{ThetavsA2}
\end{centering}
\end{figure}

To investigate the influence of lattice geometry on the magnetic properties of the {\it A}$_2$IrO$_3$ family, zero-field-cooled (ZFC) magnetization data were collected under an applied magnetic field of $\mu_0${\it H} = 1 T on each of the compounds in the Na$_{3-\delta}$(Na$_{1-x}$Mg$_x)$Ir$_2$O$_6$ series. Fig.~\ref{CWanalysis} shows plots of inverse magnetic susceptibility ($\chi^{-1}~\approx~\frac{H}{M}$) versus temperature for all compounds studied in this work. Datasets are staggered along the y-axis for visibility. Linear fits to the data in the paramagnetic regime ({\it T}~$>$~60 K) yield the Curie constant {\it C}, the Weiss temperature $\theta_W$, and the temperature-independent susceptibility $\chi_0$ (Table \ref{susceptibilitytable}). $\theta_W$, a measure of the average magnetic interaction strength, is negative for all members of the series, indicating that antiferromagnetic interactions dominate in these materials. Fig.~\ref{ThetavsA2} shows a plot of $\theta_W$ versus the average nearest neighbor Ir--Ir distance for all members of the Na$_{3-\delta}$(Na$_{1-x}$Mg$_x$)Ir$_2$O$_6$ series. There is a clear trend in the magnitude of $\theta_W$ as a function of Ir--Ir distance, illustrated with a solid red line. Contraction of the honeycomb plane results in a decrease in the magnitude of {\it $\theta_W$}. The change in interaction strength is likely due to modulation of the Ir--O--Ir bond angles, which become narrower as the lattice contracts. Contraction of the lattice causes these angles to decrease, resulting in increased ferromagnetic exchange and a concomitant decrease in the magnitude of $\theta_W$. To illustrate this effect, the average Ir--O--Ir bond angles at {\it T} = 5 K are given on the plot for the two samples that were analyzed via NPD, along with the average of the two angles reported for Na$_2$IrO$_3$ \cite{Na2IrO3revisedstructure}.

The Curie constant {\it C} is directly related to the number of Ir$^{4+}$ species in the sample, as Ir$^{3+}$, Na, and Mg are all diamagnetic. We therefore use both the nominal compositions and the calculated values for {\it C} to estimate the ratio of Ir$^{3+}$ to Ir$^{4+}$, using the compounds whose compositions are known (Na$_2$IrO$_3$,  Na$_{3}$Mg$_{0.5}$Ir$_2$O$_6$ and Na$_{2.3}$MgIr$_2$O$_6$) as reference points (Table~\ref{susceptibilitytable}). It must be noted that in the strong spin-orbit coupling limit there is a substantial orbital contribution to the observed susceptibility of the Ir$^{4+}$ electrons. Nevertheless, these estimates show unambiguously that a significant quantity of low-spin Ir$^{3+}$ can exist as non-magnetic ``holes" in {\it A}$_2$IrO$_3$ honeycomb lattices. This is unsurprising from a chemical standpoint, as the Ir$^{3+}$/Ir$^{4+}$ redox cycle is well documented in the literature, and has even been studied for use in water oxidation catalysis \cite{IrCatalysis}. It is likely that the reduced Ir species are distributed in a disordered manner throughout the lattice, which would explain the origin of the spin glass-like behavior observed in many low quality Na$_2$IrO$_3$ samples.

\begin{figure}
\begin{centering}
\includegraphics[width=3.5in]{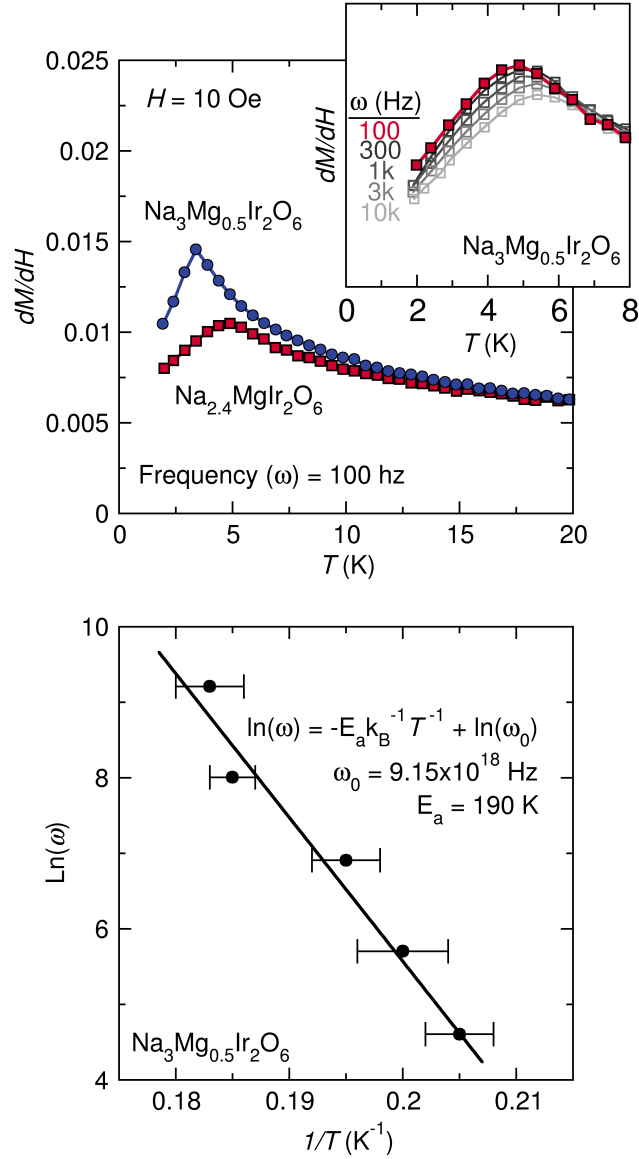}
\caption{{\bf (a)} AC susceptibility data collected on two representative samples from the Na$_{3-\delta}$Na$_{1-x}$Mg$_x$Ir$_2$O$_6$ series under an applied field of {\it H} = 796 A/m (10 Oe) with amplitude of {\it H} = 398 A/m (5 Oe). Both samples show a peak in the real component $\chi$' of the AC susceptibility at the freezing temperature {\it T$_f$}, and the magnitude of the peak decreases and shifts to higher temperatures as the frequency of the AC field is increased, consistent with a spin glass-like transition (inset). {\bf (b)} A plot of $\ln$($\omega$) versus $\frac{1}{{\it T_{f}}}$ yields a roughly linear dependence, indicating proximity to a super paramagnetic regime, and the slope and intercept of the linear fit yield activation energy {\it E$_a$} and the characteristic frequency {\it $\omega_0$} for spin reorientation.}
\label{SGanalysis}
\end{centering}
\end{figure}

\begin{figure}
\begin{centering}
\includegraphics[width=3.5in]{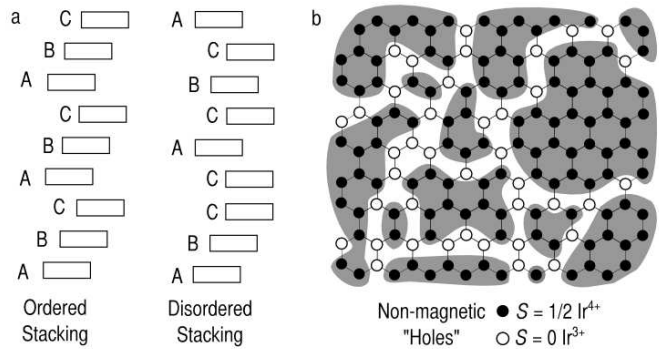}
\caption{Common types of disorder in {\it A}$_2$IrO$_3$ honeycomb iridates. {\bf (a)} In a fully ordered sample, honeycomb layers are stacked in a perfectly repeating pattern (ABCABCABC, for example). Stacking faults (ABCCACBCA, here) complicate structural characterization as they masquerade as (Na/Mg) mixing onto the Ir site. {\bf (b)} Non-magnetic Ir$^{3+}$ also exist due to chemical defects. These ``holes" perturb the magnetic order in {\it A}$_2$IrO$_3$ materials, and can lead to the disordered freezing of spins commonly observed in polycrystalline samples. We speculate that this type of disorder leads to the formation of isolated ``islands" of AFM order, which vary in size and interact weakly with one another, which is one possible explanation for the pseudo--superparamagnetic behavior observed in AC magnetic susceptibility measurements.}
\label{disorder}
\end{centering}
\end{figure}

Samples in the Na$_{3-\delta}$(Na$_{1-x}$Mg$_x$)Ir$_2$O$_6$ series exhibit spin glass-like behavior. Fig.~\ref{SGanalysis} (a) shows AC magnetic susceptibility data collected on Na$_{3}$Mg$_{0.5}$Ir$_2$O$_6$ and Na$_{2.3}$MgIr$_2$O$_6$. In both sets of data there is a broad peak in the real component of the AC susceptibility$\chi'$ at low temperatures, the position {\it T$_{f}$} and height {\it $\chi'_{max}$} of which vary as a function of the frequency $\omega$ of the AC field. Analysis of the data in the context of a canonical spin glass using the Vogel--Fulcher law or standard theory for dynamical scaling yield unphysical parameters and therefore little insight \cite{Mydosh}. The linear relationship between of $\ln$({\it $\omega$}) and $\frac{1}{{\it T_{f}}}$ indicates an Arrhenius--type activation barrier, with activation energies {\it E$_a$} between 100 K and 300 K and characteristic frequencies {\it $\omega_0$} on the scale of 10$^{15}$ Hz to 10$^{25}$ Hz \cite{Young} (Fig.~\ref{SGanalysis} (b)). While these large values, located in Table~\ref{SGtable}, are clearly unphysical, they do imply that the frozen, disordered magnetic state observed in these and other iridates is closer to superparamagnetism than to spin glass. This implication is best understood by considering the two major sources of disorder in honeycomb iridates: stacking faults (Fig.~\ref{disorder}(a)) and non-magnetic Ir$^{3+}$ ``holes"  (Fig.~\ref{disorder}(b)). Stacking faults occur due to the negligible energy difference between all possible stacking arrangements, but may have little influence on magnetic order because exchange interactions between neighboring layers are small compared to the interactions within each two dimensional layer. Non-magnetic Ir$^{3+}$ ``holes" are also likely to exist due to the ease with which Ir$^{4+}$ can be reduced. This type of defect, in contrast to stacking faults, can have a significant effect on magnetic order even at low levels, as it gives rise to the possibility of isolated antiferromagnetic domains \cite{Andrade}. Such a picture could explain the apparent proximity to a superparamagnetic state which in fact arises due to a distribution of local AF domain sizes.

\begin{figure}
\begin{centering}
\includegraphics[width=5.5in]{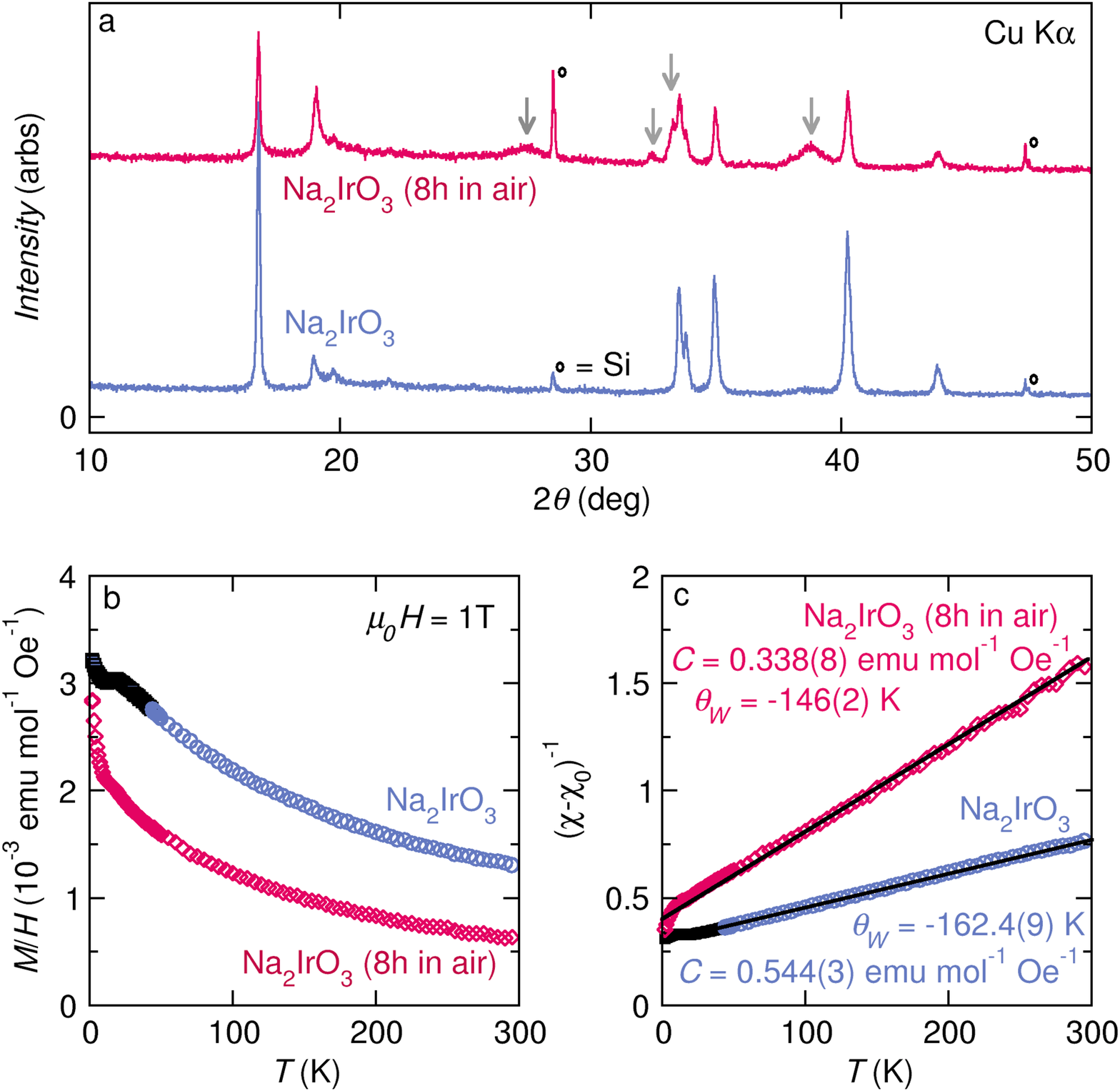}
\caption{{\bf (a)} XRPD data collected over the 2$\theta$ range 10$^{\circ}$ to 50$^{\circ}$ in six minutes on Na$_2$IrO$_3$ immediately after removal from an argon-filled glovebox (blue, bottom) and after eight hours of exposure to laboratory air (red, top). A reaction occurs between Na$_2$IrO$_3$ and one or more components of the air that causes the relative intensities of the C2/{\it m} reflections to change, and new reflections emerge, highlighted with gray arrows. Small circles indicate reflections due to crystalline Silicon, which was used as an internal standard for the purpose of Rietveld refinement. {\bf (b)} The magnetic susceptibilities of polycrystalline Na$_2$IrO$_3$ (blue circles) and the same powder after eight hours of exposure to air at room temperature (red diamonds) are compared. When handled in air-free conditions, Na$_2$IrO$_3$ exhibits long-range AFM order, as evidenced by the local maximum in the magnetic susceptibility at $T_N$ = 15 K. {\bf (c)} A plot of inverse magnetic susceptibility versus temperature, along with linear fits to both datasets above {\it T} = 60 K yield the Weiss temperature $\theta_W$ and Curie constant {\it C} for Na$_2$IrO$_3$ before (blue circles) and after (red diamonds) exposure to air.}
\label{hydration}
\end{centering}
\end{figure}

Fig.~\ref{hydration} (a) shows XRPD data collected over the 2$\theta$ range 10$^{\circ}$ to 50$^{\circ}$ in six minutes on a nearly pristine sample of Na$_2$IrO$_3$, collected immediately upon removal from an argon-filled glovebox, where the sample was ground and prepared after heating at 900 $^{\circ}$C. XRPD data collected on powder from the same sample after eight hours of exposure to laboratory air show significant changes in the relative intensities of the Bragg peaks, as well as the development of broad new peaks, however there is only a minor change in the lattice parameters of the Na$_2$IrO$_3$ unit cell over this time period. In contrast, the magnetic properties are significantly affected by the reaction. When handled properly, polycrystalline Na$_2$IrO$_3$ exhibits long range AFM order, as evidenced by the local peak in the magnetic susceptibility data below $T_N$ = 15 K (Fig.~\ref{hydration}(b)). After eight hours of exposure to air, there is no such local maximum. Curie-Weiss fitting to the linear region of the data above {\it T} = 60 K shows that $\theta_W$ decreases substantially as a result of the reaction (Fig.~\ref{hydration}(c)). The Curie constant also decreases significantly, suggesting that the average oxidation state of Ir is changing as a result of the reaction, resulting in fewer {\it S} = $\frac{1}{2}$ Ir$^{4+}$ species. The fact that the lattice parameters of the Na$_2$IrO$_3$ cell do not change significantly as a result of exposure to air suggests that this is not simply a result of hydration, as one might expect from experience with structurally analogous systems \cite{NaxCoO2}. The appearance of new peaks suggests that a chemical reaction is occurring between one or more components of laboratory air and Na$_2$IrO$_3$, a possibility that was very recently explored in detail \cite{Na2IrO3hydration}. This result suggests the origin of the discrepancy between magnetic data collected on single crystalline and polycrystalline samples: the small surface area to volume ratio of single crystalline samples hinders the reaction, leaving much of the sample's bulk unreacted. Unfortunately, this implies that previous studies on the magnetic properties and structure of single-crystalline Na$_2$IrO$_3$ may have yielded inaccurate data due to sample inhomogeneity.

\section{Conclusions}

A new series of compounds based on the {\it A}$_2$IrO$_3$ prototype with the formula Na$_{3-\delta}$(Na$_{1-x}$Mg$_x$)Ir$_2$O$_6$ (0~$\leq$~{\it x}~$\leq$~1) were synthesized and characterized using NPD, XRPD, ICP-OES, and magnetometry. Substitution of Mg for Na results in contraction of the {\it a}- and {\it b}-axes and expansion of the {\it c}-axis, suggesting that Mg substitution occurs primarily in the honeycomb plane. Similarly, Na deficiency between honeycomb planes contracts the {\it a}- and {\it b}-axes and expands the {\it c}-axis. All samples exhibit stacking disorder, which complicates structural characterization. Magnetic data collected on the compounds studied in this series suggest that there are two main variables that determine the magnetic properties of compounds in the {\it A}$_2$IrO$_3$ family: (1) the strength of the magnetic superexchange interactions between neighboring {\it S}~=~$\frac{1}{2}$ Ir$^{4+}$ species is dependent on the angle of the Ir--O--Ir bonds, for which the in plane lattice parameters {\it a} and {\it b} are a good structural marker, and (2) the number of {\it S}~=~$\frac{1}{2}$ Ir$^{4+}$ sites is influenced by the chemical composition of the compound because the oxidation state of Ir is variable. This demonstrates that chemical defects can lead to significant numbers of non-magnetic ``holes", which are likely the root of the pseudo-superparamagnetic behavior observed in many polycrystalline samples. We further showed that Na$_2$IrO$_3$ reacts quickly with laboratory air, producing significant changes in its magnetic behavior, and reported magnetic data on high-quality polycrystalline Na$_2$IrO$_3$, which shows long-range AFM order. Together, our results demonstrate that defects and disorder have significant effects of the magnetism of the {\it A}$_2$IrO$_3$ family, and that these materials can be chemically tuned to in order to explore experimentally what may prove to be a rich magnetic phase diagram.

\section{Acknowledgements}

Acknowledgement is made to the donors of the American Chemical Society Petroleum Research Fund and the David and Lucille Packard Foundation for support of this research.

\section{Tables and Figures}
\clearpage

\begin{table}
\begin{tabular}{cccccc}
 Na$_{3}$Mg$_{0.5}$Ir$_2$O$_6$ & ({\it C2}/{\it m} (12)) & & 5 K & 100 K & 295 K \\
\hline
 & & {\it a} (\AA) & 5.3565(3) & 5.3585(3) & 5.3621(5) \\
 & & {\it b} (\AA) & 9.3172(5)& 9.3216(5) & 9.3294(9) \\
 & & {\it c} (\AA) & 5.6032(3) &  5.6069(4) & 5.6186(4) \\
 & & $\beta $ ($^{\circ}$) & 108.617(7) & 108.632(8) & 108.58(1) \\
Na1 (2{\it d}) & $\frac{1}{2}$,0,$\frac{1}{2}$ & occ. & 0.81(1) & 0.81(1) & 0.81(1) \\
 & & U$_{iso}$ & 0.020(1) &  0.021(1) & 0.025(2) \\
Na2 (4{\it h}) & $\frac{1}{2}$,{\it y},$\frac{1}{2}$ & occ. & 0.81(1) & 0.81(1) & 0.81(1) \\
 & & {\it y} & 0.328(2) & 0.334(2) & 0.326(3) \\
 & & U$_{iso}$ & 0.020(1) & 0.021(1) & 0.025(2) \\
Na3/Mg1 (2{\it a}) & 0,0,0 & occ. & 0.54/0.46 & 0.54/0.46 & 0.54/0.46 \\
 & & U$_{iso}$ & 0.0247(4) & 0.0209(1) & 0.0180(5) \\
Ir1 (4{\it g}) & $\frac{1}{2}$,{\it y},0 & {\it y} & 0.1691(8) & 0.1660(6) & 0.1740(7) \\
 & & U$_{iso}$ & 0.0247(4) & 0.0209(1) & 0.0180(5) \\
O1 (8{\it j}) & {\it x},{\it y},{\it z} & {\it x} & 0.772(1) & 0.754(2) & 0.765(2) \\
 & & {\it y} & 0.1657(9) & 0.1723(7) & 0.1668(9) \\
 & & {\it z} & 0.7939(6) & 0.7956(8) & 0.794(1) \\
 & & U$_{iso}$ & 0.0197(3) & 0.0212(6) & 0.0296(9) \\
O2 (4{\it i}) & {\it x},0,{\it z} & {\it x} & 0.717(2) & 0.728(2) & 0.719(3) \\
 & & {\it z} & 0.203(1) & 0.204(2) & 0.201(2) \\
 & & U$_{iso}$ & 0.0197(3) & 0.0212(6) & 0.0296(9) \\
 Ir--O--Ir 1($^{\circ}$) & & & 99.1(4) & 97.5(3) & 100.7(6) \\
 Ir--O--Ir 2($^{\circ}$) & & & 94.0(3) & 95.0(3) & 93.8(3) \\
 \hline
 & & $\chi^2$ & 1.25 & 1.211 & 1.009 \\
 & & {\it $R_{wp}$} & 9.98 & 10.35 & 12.87 \\
 & & {\it R$_p$} & 7.95 & 8.25 & 11.09 \\
 \hline
\end{tabular}
\caption{Results of Rietveld refinement to NPD data collected on Na$_{3}$Mg$_{0.5}$Ir$_2$O$_6$ (sample 7).}
\label{NPDtable1}
\end{table}
   
\begin{table}
\begin{tabular}{cccccc}
 Na$_{2.4}$MgIr$_2$O$_6$ & ({\it C2}/{\it m} (12)) & & 5 K & 100 K \\
\hline
& & {\it a} (\AA) & 5.3084(6) & 5.3091(5) \\
& & {\it b} (\AA) & 9.197(1) & 9.1992(9) \\
& & {\it c} (\AA) & 5.6461(3) & 5.6473(3) \\
& & $\beta $ ($^{\circ}$) & 108.440(7) & 108.441(6) \\
Na1 (2{\it d}) & $\frac{1}{2}$,0,$\frac{1}{2}$ & occ. & 0.77(1) & 0.77(1) \\
 & & U$_{iso}$ & 0.025(1) &  0.041(9) \\
Na2 (4{\it h}) & $\frac{1}{2}$,{\it y},$\frac{1}{2}$ & occ. & 0.77(1) & 0.77(1) \\
 & & {\it y} & 0.331(3) & 0.343(2) \\
 & & U$_{iso}$ & 0.025(1) & 0.041(9) \\
Na3/Mg1 (2{\it a}) & 0,0,0 & occ. & 0/1 & 0/1 \\
 & & U$_{iso}$ & 0.0220(5) &  0.041(5) \\
Ir1 (4{\it g}) & $\frac{1}{2}$,{\it y},0 & {\it y} & 0.1634(9) & 0.171(1) \\
 & & U$_{iso}$ & 0.0230(5) &  0.041(5) \\
O1 (8{\it j}) & {\it x},{\it y},{\it z} & {\it x} & 0.763(2) & 0.766(2) \\
 & & {\it y} & 0.162(1) & 0.164(1) \\
 & & {\it z} & 0.802(1) & 0.798(1) \\
 & & U$_{iso}$ & 0.0184(4) & 0.012(4) \\
O2 (4{\it i}) & {\it x},0,{\it z} & {\it x} & 0.724(3) & 0.726(2)\\
& & {\it z} & 0.198(2) & 0.189(2) \\
& & U$_{iso}$ & 0.0184(4) &  0.012(4) \\
& & Ir--O--Ir 1 ($^{\circ}$) & 97.8(6) & 99.5(5) \\
& & Ir--O--Ir 2 ($^{\circ}$) & 94.4(4) & 93.9(3) \\
 \hline
 & & $\chi^2$ & 1.051 & 1.015 \\
 & & {\it $R_{wp}$} & 9.26 & 9.10 \\
 & & {\it R$_p$} & 7.95 & 8.00 \\
\hline
\end{tabular}
\caption{Results of Rietveld refinement to NPD data collected on Na$_{2.4}$MgIr$_2$O$_6$ (sample 11).}
\label{NPDtable2}
\end{table}

\begin{table}
\begin{tabular}{cccc}
 {\it Sample No.} & {\it NPD Formula} & & Na : Mg : Ir molar ratio \\
\hline
7 & Na$_{2.5}$Na$_{0.5}$Mg$_{0.5}$Ir$_2$O$_6$ & {\it ICP-OES} & 6.52 : 1.00 : 3.23 \\
 & & {\it NPD} &  7.00 : 1.00 : 4.00 \\
11 & Na$_{2.4}$MgIr$_2$O$_6$ & {\it ICP-OES} & 2.30 : 1.00 : 1.49 \\
& & {\it NPD} & 2.40 : 1.00 : 2.00\\
\hline
\end{tabular}
\caption{results of ICP-OES elemental analysis performed on Na$_3$Mg$_{0.5}$Ir$_2$O$_6$ and Na$_{2.4}$MgIr$_2$IrO$_6$ compared with compositions determined from Rietveld refinement to NPD data. Ir contents determined from ICP-OES are anomalously low due to complications which arose in the dissolution process.}
\label{ICPOES}
\end{table}

\begin{table}
\begin{tabular}{ccccccc}
{\it Sample No.} & {\it Target Stoichiometry} & {\it a} (\AA) & {\it b} (\AA) & {\it c} (\AA) & $\beta $ ($^{\circ}$) & {\it \% Disorder} \\
 \hline
1 & Na$_3$NaIr$_2$O$_6$ & 5.4288(1) & 9.4029(1) & 5.6133(2) & 108.999(4) & 30\% \\
2 & Na$_3$NaIr$_2$O$_6$ (8h in air) & 5.4270(1) & 9.3999(1) & 5.6135(2) & 108.991(5) & 20\% \\
3 & Na$_3$Na$_{0.9}$Mg$_{0.1}$Ir$_2$O$_6$ & 5.4194(1) & 9.3867(1) & 5.6088(3) & 108.63(5) & 38\% \\
4 & Na$_3$Na$_{0.8}$Mg$_{0.2}$Ir$_2$O$_6$ & 5.4132(3) & 9.3759(3) & 5.6188(3) & 108.685(6) & 50\% \\
5 & Na$_{3}$Na$_{0.7}$Mg$_{0.3}$Ir$_2$O$_6$ & 5.4036(2) & 9.3593(2) & 5.6198(3) & 108.901(6) & 37\% \\
6 & Na$_{3}$Na$_{0.5}$Mg$_{0.5}$Ir2O$_6$ & 5.3893(1) & 9.3345(1) & 5.6291(3) & 108.914(5) & 32\% \\
7 & Na$_{2.5}$Na$_{0.5}$Mg$_{0.5}$Ir$_2$O$_6$* & 5.3621(5) & 9.3294(9) & 5.6186(4) & 108.58(1) &  -- \\
8 & Na$_{3}$Na$_{0.52}$Mg$_{0.48}$Ir$_2$O$_6$ & 5.3860(2) & 9.3290(2) & 5.6316(3) & 108.880(5) & 38\% \\
9 & Na$_{2.56}$Na$_{0.5}$Mg$_{0.5}$Ir$_2$O$_6$ & 5.3623(3) & 9.2878(3) & 5.6427(1) & 108.295(2) & 30\% \\
10 & Na$_{2.46}$Na$_{0.24}$Mg$_{0.76}$Ir$_2$O$_6$ & 5.3488(1) & 9.2643(1) & 5.6459(2) & 108.620(4) & 28\% \\
11 & Na$_{2.4}$MgIr$_2$O$_6$* & 5.3171(1) & 9.2094(1) & 5.6707(3) & 108.499(6) & 32\% \\
 \hline
\end{tabular}
\caption{Target stoichiometries and unit cell parameters determined from XRPD data collected at room temperature of all compounds studied in this work. Compositions marked with an asterisk were determined by Rietveld refinement to NPD data collected on the final product and verified via ICP-OES. The 4{\it g}/2{\it a} site mixing percentages, referred to as ``\% Disorder," are a rough measure of the amount of structural disorder present in the sample.}
\label{compoundstable}
\end{table}
 
\begin{table}
\begin{tabular}{ccccccc}
{\it Sample No.} & {\it Target Stoichiometry} & $\chi_0$ ($\frac{emu}{mol Ir \cdot K}$) & {\it C} ($\frac{emu \cdot K}{mol Ir \cdot Oe}$) & $\theta_W$ (K) & Ir$^{3+}$:Ir$^{4+}$ ($\pm$0.1) \\
 \hline
1 & Na$_3$NaIr$_2$O$_6$ & 1.1(6) $\cdot$10$^{-4}$ & 0.544(3) & -162.4(9) & 0:1\\
2 & Na$_3$NaIr$_2$O$_6$(8h in air) & -5(2) $\cdot$10$^{-7}$ & 0.338(8) & -146(2) & 0.3:0.7 & \\
3 & Na$_3$Na$_{0.9}$Mg$_{0.1}$Ir$_2$O$_6$ & -9(1) $\cdot$10$^{-5}$& 0.406(7) & -87(1) & \\
4 & Na$_3$Na$_{0.8}$Mg$_{0.2}$Ir$_2$O$_6$ & -9(1) $\cdot$10$^{-5}$ & 0.345(4) & -63(2) & \\
5 & Na$_{3.06}$Na$_{0.7}$Mg$_{0.3}$Ir$_2$O$_6$ & -6(1) $\cdot$10$^{-5}$ & 0.33(1) & -48(3) &  \\
6 & Na$_{3.26}$Na$_{0.5}$Mg$_{0.5}$Ir2O$_6$ & 3(1) $\cdot$10$^{-5}$ & 0.330(5) & -35(2) & 0.4:0.6  \\
7 & Na$_{2.5}$Na$_{0.5}$Mg$_{0.5}$Ir$_2$O$_6$* & 3.7(8) $\cdot$10$^{-4}$ & 0.384(8) & -30(1) & 0.1:0.9 \\
8 & Na$_{3}$Na$_{0.52}$Mg$_{0.48}$Ir$_2$O$_6$ & -3.4(6) $\cdot$10$^{-5}$ & 0.333(6) & -28(2) & 0.3:0.7 \\
9 & Na$_{2.56}$Na$_{0.5}$Mg$_{0.5}$Ir$_2$O$_6$ & -9(1) $\cdot$10$^{-5}$ & 0.380(4) & -16(2) & 0.1:0.9 \\
10 & Na$_{2.46}$Na$_{0.24}$Mg$_{0.76}$Ir$_2$O$_6$& -1.2(9) $\cdot$10$^{-4}$ & 0.383(9) & -11(1) & \\
11 & Na$_{2.4}$MgIr$_2$O$_6$* & -1.1(7) $\cdot$10$^{-4}$ & 0.396(5) & -1(1) & 0.2:0.8 \\
 \hline
 \end{tabular}
 \caption{Curie-Weiss parameters determined from linear least-squares fitting to inverse magnetic susceptibility data in the paramagnetic regime.}
 \label{susceptibilitytable}
 \end{table}
 
 \begin{table}
 \begin{tabular}{ccccc}
 {\it Sample No.} & {\it Target Stoichiometry} & {\it T$_f$}(100 Hz) & {\it E$_a$} (K) & {\it $\omega_0$} (Hz) \\
 \hline
 5 & Na$_{3}$Na$_{0.7}$Mg$_{0.3}$Ir$_2$O$_6$ & 3.8 & 118(9) & 10$^{15}$ \\
 6 & Na$_{3}$Na$_{0.5}$Mg$_{0.5}$Ir2O$_6$ & 4.8 & 235(9) & 10$^{23}$ \\
 7 & Na$_{2.5}$Na$_{0.5}$Mg$_{0.5}$Ir$_2$O$_6$* & 4.9 & 190(13) & 10$^{18}$ \\
 8 & Na$_{3}$Na$_{0.52}$Mg$_{0.48}$Ir$_2$O$_6$ & 4.3 & 155(20) & 10$^{17}$ \\
 9 & Na$_{2.56}$Na$_{0.5}$Mg$_{0.5}$Ir$_2$O$_6$ & 4.1 & 166(14) & 10$^{19}$ \\
11 & Na$_{2.4}$MgIr$_2$O$_6$* & 3.4 & 187(11) & 10$^{25}$ \\
 \hline
 \end{tabular}
 \caption{Freezing temperatures {\it T$_f$}, activation energies {\it E$_a$}, and characteristic frequencies {\it $\omega_0$} determined from an Arrhenius analysis of AC magnetic susceptibility data collected on members of the Na$_{3-\delta}$(Na$_{1-x}$Mg$_x$)Ir$_2$O$_6$ series.}
 \label{SGtable}
 \end{table}

\begin{thebibliography}{99}

\bibitem{MottStrongSOC}G. Jackeli \& G. Khaliullin, Mott Insulators in the Strong Spin-Orbit Coupling Limit: From Heisenberg to a Quantum Compass and Kitaev Models. {\it Physical Review Letters} {\bf 102,} 017205 (2009)

\bibitem{Na2IrO3Prediction}A. Shitade, H. Katsura, J. Kunes, X.-L. Qi, S.-C. Zhang, N. Nagaosa, Quantum Spin Hall Effect in a Transition Metal Oxide Na$_2$IrO$_3$. {\it Physical Review Letters} {\bf 102,} 256401 (2009)

\bibitem{XtalfieldandcorrelationA2IrO3}H. Gretarsson et al., Crystal-Field Splitting and Correlation Effect on the electronic structure of {\it A}$_2$IrO$_3$. {\it Physical Review Letters} {\bf 110,} 076402 (2013)

\bibitem{A2IrO3KitaevHeisenberg}J. Chaloupka, G.  Jackeli \& G. Khaliullin, Kitaev-Heisenberg Model on a Honeycomb Lattice: Possible Exotic Phases in Iridium Oxides {\it A}$_2$IrO$_3$. {\it Physical Review Letters} {\bf 105,} 027204 (2010)

\bibitem{Kimchi}I. Kimchi, J. J. Analytis, A. Vishwanath, Three dimensional quantum spin liquid in a hyperhoneycomb iridate model and phase diagram in an infinite--D approximation. arXiv:1309.1171 (2013)

\bibitem{Andrade}E. C. Andrade, M. Vojta, Magnetism in spin models for depleted honeycomb--lattice iridates: Spin--glass order towards percolation. arXiv:1309.2951 (2013)

\bibitem{Rau}J. G. Rau, E. K.-H. Lee, H.-Y. Kee, Generic spin model for the honeycomb iridates beyond the Kitaev limit. arXiv:1310.7940 (2013)

\bibitem{Lei}H. Lei, Z. Zhong, H. Hosono, Structural, Magnetic, and Electrical Properties of Li$_2$Ir$_{1-x}$Ru$_x$O$_3$. arXiv:1311.7317 (2013)

\bibitem{Katukuri}V. M. Katukuri, {\it et al.}, Kitaev interactions between {\it j}~=~1/2 moments in honeycomb Na$_2$IrO$_3$ are large and ferromagnetic: Insights from {\it ab initio} quantum chemistry calculations. arXiv:1312.7437 (2013)

\bibitem{Na2IrO3Synthesis}Y.Singh \& P. Gegenwart, Antiferromagnetic Mott insulating state in single crystals of the honeycomb lattice material Na$_2$IrO$_3$. {\it Physical Review B} {\bf 82}, 064412 (2010)

\bibitem{Na2IrO3revisedstructure}S. K. Choi {\it et al.}, Spin Waves and Revised Crystal Structure of Honeycomb Iridate Na$_2$IrO$_2$. {\it Physical Review Letters} {\bf 108}, 127204 (2012)

\bibitem{NaLiDoping}
G. Cao {\it et al.}, Evolution of Magnetism in Single-Crystal Honeycomb Iridates. arXiv:1307.2212v1 (2013)

\bibitem{BaroudiCava}K. Baroudi {\it et al.}, Structure and Properties of $\alpha$-NaFeO$_2$-type Ternary Sodium Iridates. arXiv:1312.0995 (2013)

\bibitem{Na2IrO3hydration}J. W. Krizan, J. H. Roudebush, G. M. Fox, R. J. Cava., The Chemical Instability of Na$_2$IrO$_3$ in Air. arXiv:1312.1637 (2013)

\bibitem{Li2MnO3}J. Br\'eger {\it et al.}, High-resolution X-ray diffraction, DIFFaX, NMR and first principles study of disorder in the Li$_2$MnO$_3$--Li[Ni$_{1/2}$Mn$_{1/2}$]O$_2$ solid solution. {\it Journal of Solid State Chemistry}, {\bf 178}, 2575 (2005) 

\bibitem{GSAS}A. C. Larson, R. B. Von Dreele, {\it Los Alamos National Laboratory Report LAUR}, 86--748. (2000)

\bibitem{EXPGUI}B. H. Toby, EXPGUI, a graphical user interface for GSAS, {\it J. Appl. Crystallogr.}, {\bf 34}, 210--213. (2001)

\bibitem{DIFFaX}M. M. J. Treacy, J. M. Newsam \& M. W. Deem, A general recursion method for calculating diffracted intensities from crystals containing planar faults. {\it Proceedings of the Royal Society, London} {\bf A433}, 499-520 (1991)

\bibitem{Cu5SbO6}E. Climente-Pascual, {\it et. al}, Spin $\frac{1}{2}$ Delafossite Honeycomb Compound Cu$_5$SbO$_6$, {\it Journal of Inorganic Chemistry}, {\bf 51}, 557 (2012)

\bibitem{IrCatalysis}J. D. Bakemore, {\it et al}, Half-Sandwich Iridium Complexes for Homogeneous Water-Oxidation Catalysis. {\it Journal of the American Chemical Society}, {\bf 132}, 16017 (2010)

\bibitem{Mydosh}J. A. Mydosh, Spin Glasses: An Experimental Introduction, Taylor and Francis Inc. (1993) ISBN 0-7484-0038-9

\bibitem{Young}K. Binder, A. P. Young, Spin glasses: Experimental facts, theoretical concepts, and open questions. {\it Rev. Mod. Phys.}, {\bf 58}, 801 (1986)

\bibitem{NaxCoO2}J. W. Lynn {\it et al.}, Structure and dynamics of superconducting Na$_x$CoO$_2$ hydrate and its unhydrated analog., {\it Physical Review B}, {\bf 68}, 214516 (2003)

\end{thebibliography}
\end{document}